# Symmetrical broken and nonlinear response of Weyl semimetal TaAs influenced by the topological surface states and Weyl nodes

*Shumeng Chi[1,3], Zhilin Li[2,3], Haohai Yu[1*], Gang Wang[2**], Shuxian Wang[1], Huaijin Zhang[1***] & Jiyang Wang[1]*

[1]State Key Laboratory of Crystal Materials and Institute of Crystal Materials, Shandong University, Jinan 250100, China.

[2]Beijing National Laboratory for Condensed Matter Physics, Institute of Physics, Chinese Academy of Sciences, Beijing 100190, China.

[3]S. Chi and Z. Li contributed equally to this work.

*and ** Correspondence: haohaiyu@sdu.edu.cn, and gangwang@iphy.ac.cn

# Abstract

A Weyl semimetal (WSM) features Weyl fermions in its bulk and topological surface states on surfaces, and is novel material hosting Weyl fermions, a kind of fundamental particles. The WSM was regarded as a three-dimensional version of "graphene" under the illusion. In order to explore its promising photoelectric properties and applications in photonics and photoelectronics, here, we study the anisotropic linear and nonlinear optical responses of a WSM TaAs, which are determined by the relationship and balance between its topological surface states and Weyl nodes. We demonstrate that topological surface states which break the bulk symmetry are responsible for the anisotropy of the mobility, and the anisotropic nonlinear response shows saturable characteristic with extremely large saturable intensity. We also find that the mobility is anisotropic with the magnitude of $10^4$ cm$^2$V$^{-1}$s$^{-1}$ at room temperature and can be accelerated by the optical field. By analyzing the symmetry, the nonlinear response is mainly contributed by the fermions close to the Weyl nodes, and is related to the Pauli's blocking of fermions, electron-electron interaction. This work experimentally discovers the anisotropic ultrahigh mobility of WSMs in the optical field and may start the field for the applications of WSMs in photonics and photoelectronics.

# INTRODUCTION

Weyl fermions are a kind of fundamental particles with definite chirality and can be described by a massless solution of the Dirac equation as proposed by Weyl in 1929 in quantum field theory[1-3]. Since then, the hunting for Weyl fermions has undergone for decades. Due to the developments in the field of topological insulators[4-6], it has been realized that the Weyl fermions can emerge in certain novel semimetals with non-trivial topology with either broken time-reversal symmetry or inversion symmetry as quasiparticle excitations [7-16]. These semimetals are so called as Weyl semimetals (WSMs). The Weyl fermions at zero energy correspond to the Weyl nodes formed by crossing of nondegenerate bulk bands in the three-dimensional (3D) momentum space. The Weyl nodes come in pairs with opposite chirality in 3D crystals due to no-go theorem[17, 18]. Therefore, WSM can not be simply regarded as a 3D version of "graphene" [3, 15]. When the bulk Weyl nodes are projected onto the surface Brillouin zone (SBZ), the projection points are connected by the Fermi arcs, which are composed by the Fermi points of gapless chiral topological surface states. It has been revealed that the Weyl nodes and surface states endue unusual magneto-transport properties to the WSM in a magnetic field under low temperature including the large, non-saturating and anisotropic magnetoresistance [1, 2, 19-23].

Under an optical field, the Weyl nodes and Fermi arcs interact with the photons and the study on this interaction process may discover electronic and optoelectronic properties of the WSM and inspire the promising applications in photonics and optoelectronics, such as phototransistors, photodetectors, optical modulators, etc, analogous to graphene[16]. In addition, the nonlinear response of electrons under strong light fields is the origination of nonlinear optics[17] and photonics[18]. Here, we report the anisotropic linear and nonlinear optical response of TaAs crystal. By the careful investigation of linear optical conductivity, we find that the mobility of TaAs is ultrahigh and anisotropic, which is mainly related to the reduced symmetry of the surface states ($C_4 \rightarrow C_2$). However, under the strong light field, the anisotropic nonlinear response can be saturated with extremely large saturable intensity, almost preserves the bulk $C_4$ symmetry and also positively contributes to the mobility.

## RESULTS&DISCUSSION

**Anisotropic ultrahigh mobility:**

Under a light irradiation with the frequency of ω and electric field $\widehat{E}(t) = \widehat{E}\exp(-i\omega t)$, as obtained by the Drude-Sommerfeld model[24], the generated current density $\widehat{J}$ can be shown as:

$$\widehat{J} = \sigma \widehat{E}(t) = \sigma \widehat{E}\exp(-i\omega t) \tag{1}$$

where σ is the optical conductivity. Since the electric field and current density are tensors, the optical conductivity is a second-order tensor and should be expressed as:

$$\sigma = \begin{pmatrix} \sigma_{xx} & \sigma_{xy} & \sigma_{xz} \\ \sigma_{yx} & \sigma_{yy} & \sigma_{yz} \\ \sigma_{zx} & \sigma_{zy} & \sigma_{zz} \end{pmatrix} \tag{2}$$

In the principle axes of a crystal, the optical conductivity can be rewritten as[25]:

$$\sigma = \begin{pmatrix} \sigma_{xx} & 0 & 0 \\ 0 & \sigma_{yy} & 0 \\ 0 & 0 & \sigma_{zz} \end{pmatrix} \tag{3}$$

with $\sigma_{xy}=\sigma_{xz}=\sigma_{yx}=\sigma_{yz}=\sigma_{zx}=\sigma_{zy}=0$.

In nature, the Weyl semimetal is a single crystalline material and the relationship between the distribution of its physical property and structural symmetry should obey the Neumann principle[25-27], which indicates that the distribution of the physical properties in 3D space should obey the 3D structural symmetry. Based on the Neumann principle, $\sigma_{xx}=\sigma_{yy}$ is for the uniaxial crystal and $\sigma_{xx}\neq\sigma_{yy}$ for the biaxial crystal[25-27]. Therefore, for a Weyl semimetal, whether the symmetry is broken from the uniaxial to the biaxial is determined by the optical conductivities $\sigma_{xx}$ and $\sigma_{yy}$.

TaAs was identified as a Weyl semimetal with the space group of $I4_1md$ ($C_{4v}$) and the lattice constants of $a=b=3.437$Å and $c=11.656$ Å[15] and is a uniaxial crystal with $\sigma_{xx}=\sigma_{yy}\neq\sigma_{zz}$. The bulk and surface states are schematically shown in Fig. 1. In the bulk states, the $C_4$ screw axis is along the $c$ axis. However, the Fermi arcs formed by the surface states on (001) surface distribute differently along the x and y directions corresponding to the $k_x$ and $k_y$ directions of the surface Brillouin zone, respectively, which means that in the surface states, the bulk $C_4$ screw symmetry should be broken and become $C_2$ with the screw axis along the x and y axis[3, 15, 28]. Considering the distribution of surface states and bulk states, we can get the conclusion that for the bulk states, the optical conductivity $\sigma_{xx}=\sigma_{yy}\neq\sigma_{zz}$, however, for the surface states, $\sigma_{xx}\neq\sigma_{yy}\neq\sigma_{zz}$, and the broken symmetry of the surface states can only be distinguished by the sample with the (001) plane, a surface normal to $c$ axis, since the bulk states is isotropic only in this surface. In addition, the (001) plane is an easy-cleaving plane in angle-resolved photoemission spectroscopy measurements terminated with either Ta or As atoms[28]. Therefore, a TaAs crystal with polished (001) surface was chosen for the study of the optical conductivity of surface states.

A continuous-wave laser with the wavelength of 1.03 μm was employed as the light source for the study of room-temperature optical conductivity under weak irradiation. Detailed information of the measurements is given in the Methods. For the semimetal, the refractive index can be written with real ($n_r$) and imaginary ($n_i$) parts as[29]:

$$n = n_r + i n_i \tag{4}$$

with the relation of:

$$n_r^2 - n_i^2 = 1 - \frac{\omega_p^2}{\omega^2 + \tau^{-2}}$$
$$2 n_r n_i = \frac{\sigma}{\omega \varepsilon_0} \tag{5}$$

here, $\varepsilon_0$ is vacuum dielectric constant, $\omega_p$ is the plasma frequency, $\frac{1}{\tau}$ is the scattering rate. Based on the recent reported results about the scattering rate and Drude weight of TaAs, we can find that the incident light frequency is comparable with the plasma frequency (about $7 \times 10^3$ cm$^{-1}$)[30] and in this regime, the refractive index can be assumed as: $n_r \approx n_i$ based on eq. (5). By measuring the reflectivity R under normal reflection of different polarized incident light, the anisotropic optical conductivity σ can be achieved, since the normal reflectivity R can be shown as:

$$R = \left|\frac{n-1}{n+1}\right|^2 = \frac{(1-n_r)^2 + n_i^2}{(1+n_r)^2 + n_i^2} \approx 1 - \sqrt{\frac{8\omega\varepsilon_0}{\sigma_{ij}}} \tag{6}$$

The measured reflection at different polarization directions was shown in Fig. 2a, where the angles θ=0° and θ=90° correspond to the x and y axis, respectively. From this figure, we can find that the reflectivity is almost similar with the previous reported reflectivity without considering the anisotropy[30], however, the present reflectivity is anisotropic with R=51% and 49% at θ=0° and θ=90°, respectively, which clearly exhibited that the bulk $C_4$ screw symmetry is broken by the fermion surface states. Fitting the reflectivity shown in Fig. 2a with equations (1), (3) and (6), we can get the anisotropic refractive index and optical conductivity with the results shown in Fig. 2b and Fig. 2c, respectively. From these figures, the refractive index can be fitted with the following equation[25]:

$$n_r \approx n_i = \sqrt{\frac{n_x^2 n_y^2}{n_y^2 \cos^2(\theta) + n_x^2 \sin^2(\theta)}} \qquad (7)$$

with the results of the principal refractive indexes $n_x = 3.937$ and $n_y = 3.838$, and the optical conductivity is shown as:

$$\sigma = \begin{pmatrix} \sigma_{xx} & 0 \\ 0 & \sigma_{yy} \end{pmatrix} = \begin{pmatrix} 5 & 0 \\ 0 & 4.7 \end{pmatrix} \times 10^3 \Omega^{-1} cm^{-1} \qquad (8)$$

Based on eq. (8), the theoretical distribution of the optical conductivity can be achieved and shown in Fig. 2c. It has been theoretically predicted that TaAs has 12 pairs of Weyl nodes in total classified into two types: four pairs (W1) in the $k_z=0$ plane with 2 meV above the Fermi energy, and eight pairs (W2) located off the $k_z=0$ plane with 21 meV below the Fermi energy, and Fermi arcs connect projection points of the Weyl nodes in the surface Brillouin zone[3, 4, 15]. The incident photons have much larger energy than 21 meV, which means that both of the interband transitions nearby W1 and W2 can be realized. However, the optical conductivity no matter influenced by the Weyl nodes or surface arcs should obey the Neumann principle[25-27]. In structures, TaAs has a space group of $I4_1md$ ($C_{4v}$) in the bulk and when projected in the surface Brillouin zone, the symmetry is reduced to be the $C_2$ screw. Meanwhile, the distribution of the surface arcs obeys the $C_2$ screw symmetry. Considering the weak irradiation of the incident light field, the incident photons mainly interacted with the electrons within the surface region which are localized by the surface states. Therefore, based on the results about optical conductivity shown in eq. (8), we can get the conclusion that the anisotropy of the optical conductivity under weak irradiations is mainly contributed by the surface states.

For the present conditions $\hbar\omega \gg k_B T$ with the reduced Planck's constant $\hbar = \frac{h}{2\pi}$, Boltzmann constant $k_B$, and room temperature $T$, the conductivity of per Weyl node can be expressed as $\sigma = \frac{e^2 \omega}{12 h v_F}$ [31], with the Fermi velocity $v_F$. The Fermi velocities $v_F$ can be calculated to be $1.9 \times 10^5$ cm/s and $2.0 \times 10^5$ cm/s along the $k_x$ and $k_y$ directions, respectively. For the semimetal with Weyl nodes, the mobility is expressed as $\mu = \frac{e v_F \tau_{tr}}{\hbar k_F}$, with the transport lifetime $\tau_{tr}$ and Fermi wave vector $k_F = 0.015$Å$^{-1}$ [32]. With the reported scattering rate at room temperature[30], we can get the transport time $\tau_{tr} = 470$ fs. The mobilities along $k_x$ and $k_y$ are $8.6 \times 10^3$ cm$^2$V$^{-1}$s$^{-1}$ and $9.1 \times 10^3$ cm$^2$V$^{-1}$s$^{-1}$, respectively, which are comparable with that of graphene at room temperature ($1 \times 10^4$ cm$^2$V$^{-1}$s$^{-1}$) [33]. Associated with the symmetry of the surface and bulk states, we can get the conclusion that the difference between the mobilities in the $k_x$ and $k_y$ directions should be generated by the reduced C$_2$ symmetry of surface states in agreement with the optical conductivity. Considering the high mobility of the semimetal TaAs, we propose that it should have promising applications in electronics and optoelectronics including the transistors, photodetectors and photovoltaic devices, which are dependent on the mobility[16].

**Anisotropic nonlinear optical conductivity and acceleration of mobility**

Under high-intensity irradiations, the nonlinear response of materials would be generated and is characterized by the high-order susceptibility which is associated with the polarization per unit volume to the powers of the electric filed of the light. In a conducting material, such as graphene and semimetal, it is appropriate to characterize the response including the linear and nonlinear in terms of current density $\vec{j}$ [20, 21]. Since the second power of electric field relates to the second harmonic generation which requires the "phase-matching" indicating typical incident angles and polarization directions, and the semimetal has strong and broadband absorption[17], it is impossible to detect the second harmonic generation. Associated with the higher-order nonlinear coefficient has orders-of-magnitude smaller than the lower-order one[17], the present experiments were focused on the third-order nonlinearity. The current density is expressed as [34]:

$$J_i = \sigma_{ij}E_j(t) + \sigma_{ijkl}E_j(t)E_k(t)E_l(t)$$
$$= [\sigma_{ij} + \sigma_{ijkl}E_k(t)E_l(t)]E_j(t) \quad (9)$$

where $\sigma_{ijkl}$ is the third-order nonlinear optical conductivity corresponding to the nonlinear response of the electrons, and $J_m$ and $E_m$ are the components of the current density and electric field along $m$ direction, respectively. The third-order nonlinear optical conductivity can be expressed as[27, 34]:

$$\sigma_{ijkl} = \begin{pmatrix} \sigma_{11} & \sigma_{12} & \sigma_{13} & \sigma_{14} & \sigma_{15} & \sigma_{16} \\ \sigma_{21} & \sigma_{22} & \sigma_{23} & \sigma_{24} & \sigma_{25} & \sigma_{26} \\ \sigma_{31} & \sigma_{32} & \sigma_{33} & \sigma_{34} & \sigma_{35} & \sigma_{36} \\ \sigma_{41} & \sigma_{42} & \sigma_{43} & \sigma_{44} & \sigma_{45} & \sigma_{46} \\ \sigma_{51} & \sigma_{52} & \sigma_{53} & \sigma_{54} & \sigma_{55} & \sigma_{55} \\ \sigma_{61} & \sigma_{62} & \sigma_{63} & \sigma_{64} & \sigma_{65} & \sigma_{66} \end{pmatrix} \quad (10)$$

with1=xx, 2=yy, 3=zz, 4=yz=zy, 5=xz=zx, and 6=xy=yx.

For the incident light propagating normal to the (001) surface of TaAs, the third-order nonlinear optical conductivity can be reduced as:

$$\sigma_{ijkl} = \begin{bmatrix} \sigma_{11} & \sigma_{12} & 0 \\ \sigma_{21} & \sigma_{22} & 0 \\ 0 & 0 & \sigma_{66} \end{bmatrix} \quad (11)$$

with $\sigma_{11}=\sigma_{22}$ and $\sigma_{12}=\sigma_{21}$ for the bulk states with $C_4$ screw symmetry, $\sigma_{11} \neq \sigma_{22}$ and $\sigma_{12} \neq \sigma_{21}$ for the surface states with $C_2$ screw symmetry. Without loss of generality, the nonlinear optical conductivity was investigated under the $C_2$ screw symmetry. The optical conductivity can be exhibited as:

$$\sigma = \sigma_L + \sigma_{NL}$$
$$= \begin{pmatrix} \sigma_{xx} & 0 \\ 0 & \sigma_{yy} \end{pmatrix} + \begin{bmatrix} \sigma_{11} & \sigma_{12} & 0 \\ \sigma_{21} & \sigma_{22} & 0 \\ 0 & 0 & \sigma_{66} \end{bmatrix} \begin{bmatrix} E_x^2 \\ E_y^2 \\ E_x E_y \end{bmatrix} \quad (12)$$

The reflectivity under different incident light intensity and different polarization directions is shown in Fig. 3 which exhibits anisotropic properties. With eq. (6), the optical conductivity at different polarization directions can be derived. By fitting the optical conductivity based on the light intensity:

$$I = \frac{n}{2\mu c} E^2 \quad (13)$$

where $\mu$ is the permeability of vacuum, the nonlinear optical conductivities at different incident light intensities and polarizations can be obtained with the normalized values shown in Fig. 4a-e. In order to get the different nonlinear optical conductivity tensors, we need fall back on eq. (12) with $E_x = E\cos\theta$ and $E_y = E\sin\theta$. By fitting nonlinear optical conductivities at different polarization directions, the tensors are achieved as shown in Fig. 5 and the optical conductivity can be expressed as:

$$\sigma = \begin{pmatrix} \sigma_{xx} & 0 \\ 0 & \sigma_{yy} \end{pmatrix} + \begin{bmatrix} \sigma_{11} & \sigma_{12} & 0 \\ \sigma_{21} & \sigma_{22} & 0 \\ 0 & 0 & \sigma_{66} \end{bmatrix} \begin{bmatrix} E_x^2 \\ E_y^2 \\ E_x E_y \end{bmatrix}$$

$$= \begin{pmatrix} 5 & 0 \\ 0 & 4.7 \end{pmatrix} \times 10^3 \Omega^{-1} cm^{-1} + \begin{bmatrix} -8.2 & -10.3 & 0 \\ -10.8 & -8.8 & 0 \\ 0 & 0 & -5.2 \end{bmatrix} \times 10^{-16} \Omega^{-1} V^{-2} \times \begin{bmatrix} E_x^2 \\ E_y^2 \\ E_x E_y \end{bmatrix}$$

(14)

From eq. (14), we can find the $\sigma_{11} \approx \sigma_{22}$ and $\sigma_{12} \approx \sigma_{21}$, which indicates that under high-intensity irradiations, the $C_4$ screw symmetry is almost conserved and bulk states are responsible for the nonlinear response. The minus means that absorption by the semimetal TaAs is saturable under high-intensity irradiations, since the absorption coefficient $2\alpha$ is related to the optical conductivity as[24]:

$$\alpha = \frac{\omega}{c} n_i = \frac{1}{c} \sqrt{\frac{\sigma \omega}{2\varepsilon_0}} \qquad (15)$$

By calculation, the absorption coefficient is found to be about $48 \times 10^4$ cm$^{-1}$ with the penetration depth $\delta = \frac{1}{\alpha}$ of about 42 nm, which indicates that the absorption is mainly contributed by the bulk states under high-intensity irradiation and is generated by the interband transitions of electrons close W1 and W2. The nonlinear optical process during the photon-electron interaction is shown in Fig. 6. Under the high-intensity light irradiations, the final electronic states in the valence and conduction bands would be fully occupied, which blocks the further interband transitions due to the Pauli's blocking of fermions and the absorption becomes saturable. Based on eq. (15), we can get the saturable intensity, the irradiation intensity reducing absorption coefficient to be a half, was estimated to be $10^3$ GW/cm$^2$, which is much larger than that of graphene (0.71 to 0.61 MW/cm$^2$) [35]. The larger saturable intensity is due to the large absorption coefficient and indicates that more photon-generated carriers participate into the light-matter interaction process, compared with graphene[35, 36]. The large saturable intensity also means that the linear response of fermions is stable in agreement with the results gotten with InP, another topological Weyl semimetal candidate, in strong magneto fields[1, 28]. Up to now, the saturable absorption effect has been widely used in photonics including the optical switch in ultrafast phonics essential for the generation of short and ultrafast pulsed lasers. Based on the large staurable intensity and broadband absorption properties, we also propose that the Weyl semimetal TaAs should have promising applications as an absorber in photovoltaic devices.

Considering the relations among the mobility, Fermi velocity, and optical conductivity[31, 32], the mobility can be expressed as:

$$\mu \approx \frac{e^3 \omega \tau_{tr}}{24\pi \sigma_L \hbar^2 k_F}(1-\frac{\sigma_{NL}}{\sigma_L}) \tag{16}$$

Whether the mobility is accelerated or decelerated depend on the sign of the nonlinear optical conductivity. For the semimetal TaAs, the nonlinear optical conductivity is minus as shown in eq. (14), which indicates that the mobility is accelerated by the light filed. Based on eq. (16) and (14), the respective mobility along $k_x$ and $k_y$ can be expressed as:

$$\begin{aligned}\mu_y &= 9.1\times 10^3 \times (1+1.87\times 10^{-19} E_x^2 + 2.3\times 10^{-19} E_y^2) cm^2 V^{-1} s^{-1} \\ \mu_y &= 9.1\times 10^3 \times (1+1.87\times 10^{-19} E_x^2 + 2.3\times 10^{-19} E_y^2) cm^2 V^{-1} s^{-1}\end{aligned} \tag{17}.$$

The results indicate that the mobility of semimetal can be further accelerated by applying strong electric or light fields, which would be helpful for the modification of material mobility for the purpose of applications.

## CONCLUSION

In conclusion, the optical conductivity of a WSM TaAs has been studied including both the linear and nonlinear responses. From the optical conductivity tensors, we find that TaAs has ultrahigh mobility at room temperature comparable with that of graphene. Under high-intensity irradiations, the interband transitions nearby the W1 and W2 Weyl nodes generate the saturable absorption of light due to Pauli's blocking of fermions, which results in the acceleration of the mobility. Considering the gapless Weyl nodes, we propose that WSM should be responsible for wide-broad light and is a promising candidate being applied in electronics and photonics, including transistors, photodetectors, optical switchers, absorber in photovoltaic devices, etc.

## METHODS

TaAs growth and preparations:

TaAs single crystals were grown by the chemical vapor transport (CVT) method[37]. Tantalum foil (99.99%), arsenic pieces (99.995%), and iodine (99.99%) were filled in a silica ampule with mole ratio Ta:As:$I_2$ =1:1:0.05. Evacuated and sealed, the silica ampule was then heated gradually from room temperature to 1000 ℃ more than 72 h. TaAs polycrystalline was formed during the process. Afterward the ampule was kept in a temperature field of $\Delta T$=1020-980 ℃ for two weeks for follow-up procedure of CVT. Finally, TaAs single crystals were obtained after the ampule cooling down to room temperature naturally. The (001) surface on the as-grown TaAs single crystal is polished for the measurements of reflectivity.

Z-scan technique:

A mode-locked laser was used as the light source with the wavelength of 1.03 μm, a pulse width of 450 fs, and repetition rate of 200 kHz. The laser light was split into two beams using a beam splitter as the reference light and incident light, respectively. The incident light was focused by a convex lens with a focus length of 250 mm. The beam radius at the focus was 80 μm. The (001) surface of TaAs sample was put against the incident light with an angle less than 2° between the incident light and the normal direction of the (001) surface and could be moved along the z-axis using a linear motorized stage (NRT150/M) controlled by a computer. The optical energy from the two branches was detected simultaneously by two energy meters (S121C). Nonlinear optical conductivity was acquired by measuring the reflectivity with respect to the position z of the sample. To investigate the anisotropy of the optical conductivity, we added a half wave plate to obtain different polarized incident light.

## ACKNOWLEDGEMENT

The authors would like to thank Prof. H.M. Weng, X.L. Chen and Prof. L.W. Guo of Institute of Physics, CAS for the fruitful discussions. This work was supported by the National Natural Science Foundation of China (Grant Nos. 51422205, 5132221, 51572291), the Strategic Priority Research Program (B) of the Chinese Academy of Sciences (Grant No. XDB07020100), Natural Science Foundation for Distinguished Young Scholars of Shandong Province (JQ201415), Taishan Scholar Foundation of Shandong Province, China.

**Figure captions:**

Figure 1. Schematic bulk and surface states of WSM TaAs.

Figure 2. Anisotropic properties of WSM TaAs at room-temperature under weak irradiation. (a) The measured reflection and fitting results with eq. (6) at different polarization directions. (b) The anisotropic refractive index and fitting results with eq. (7) at different polarization directions. (c) The anisotropic optical conductivity and fitting results with eq. (8) at different polarization directions.

Figure 3. Reflectivity of TaAs surface under different incident light intensity and different polarization directions θ.

Figure 4. Normalized optical conductivity and linear fitting results based on eq. (14). (a)-(e) represent different polarization directions θ.

Figure 5. Normalized variation of optical conductivity at different polarization direction θ and theoretical fitting with eq. (14).

Figure 6. The schematically absorption process under different irradiations. (a) Linear absorption process under week irradiations corresponding to interband transition from the valence states to the conduction states. (b) Nonlinear saturable absorption under strong irradiations corresponding the fully occupied states blocking further absorption.

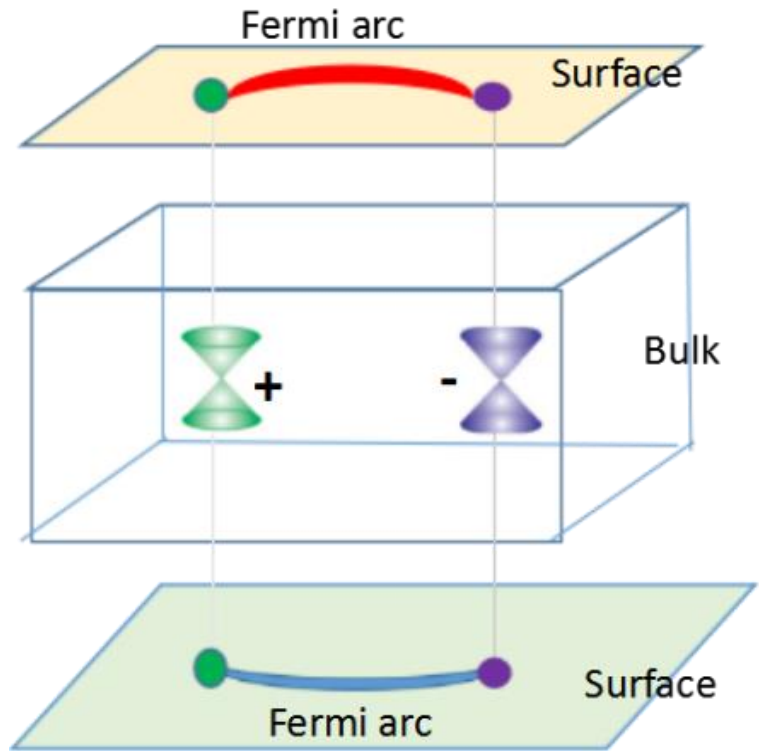

Figure 1. Schematic bulk and surface states of WSM TaAs.

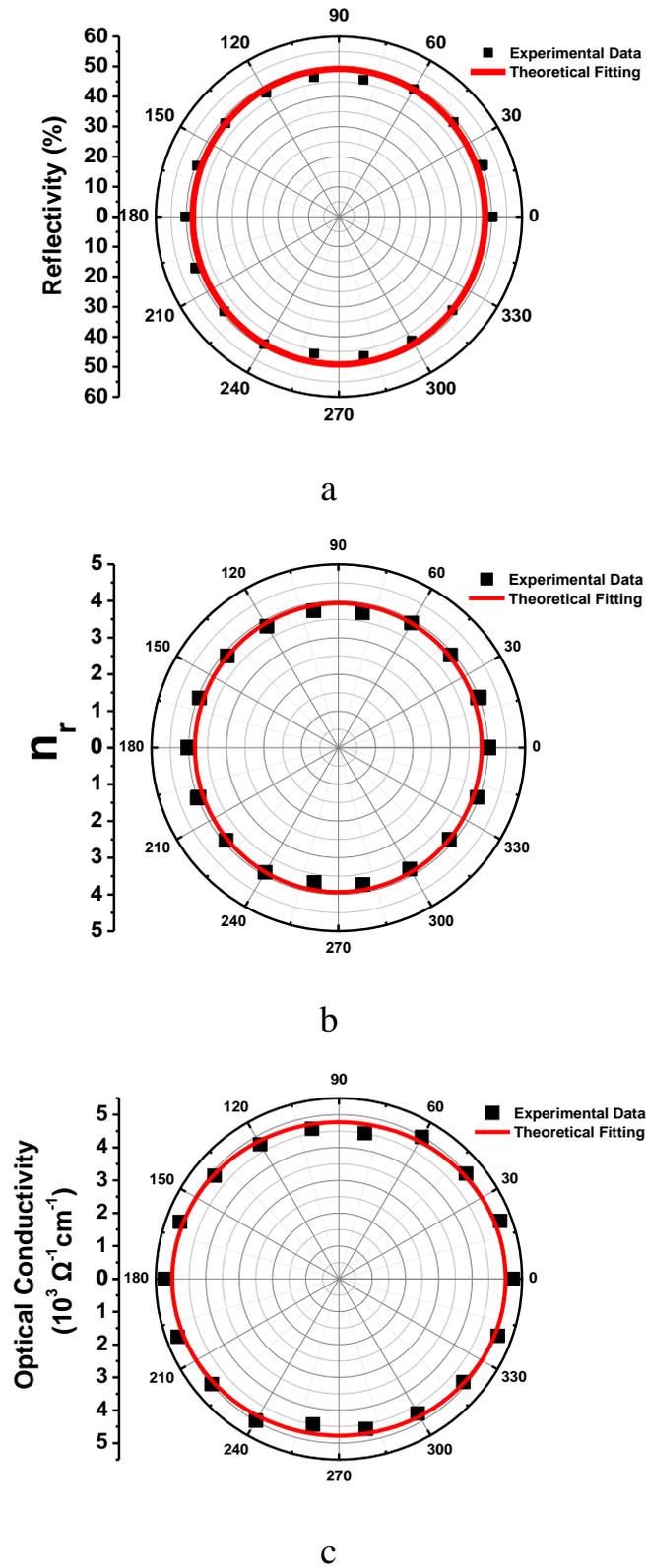

Figure 2. Anisotropic properties of WSM TaAs at room-temperature under weak irradiation. (a) The measured reflection and fitting results with eq. (6) at different polarization directions. (b) The anisotropic refractive index and fitting results with eq. (7) at different polarization directions. (c) The anisotropic optical conductivity and fitting results with eq. (8) at different polarization directions.

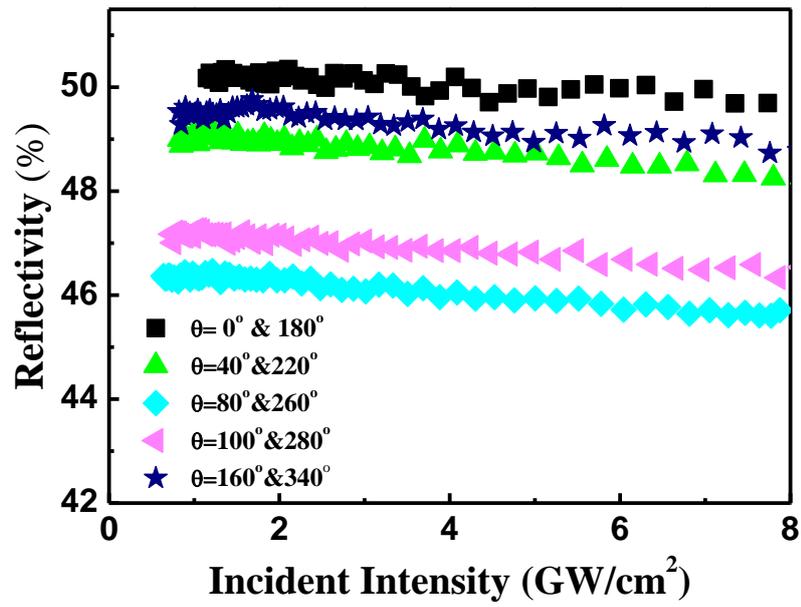

Figure 3. Reflectivity of TaAs surface under different incident light intensity and different polarization directions θ.

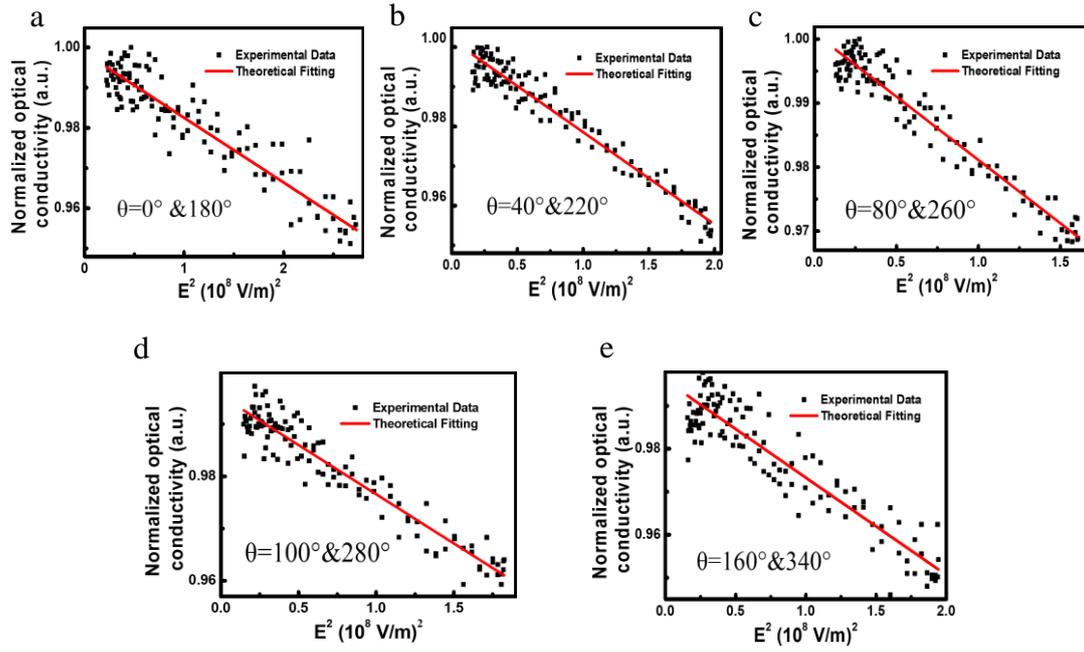

Figure 4. Normalized optical conductivity and linear fitting results based on eq. (14). (a)-(e) represent different polarization directions θ.

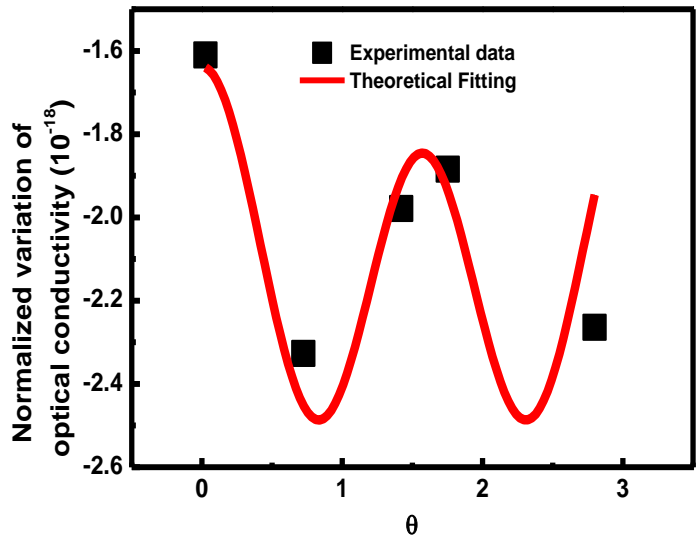

Figure 5. Normalized variation of optical conductivity at different polarization direction θ and theoretical fitting with eq. (14).

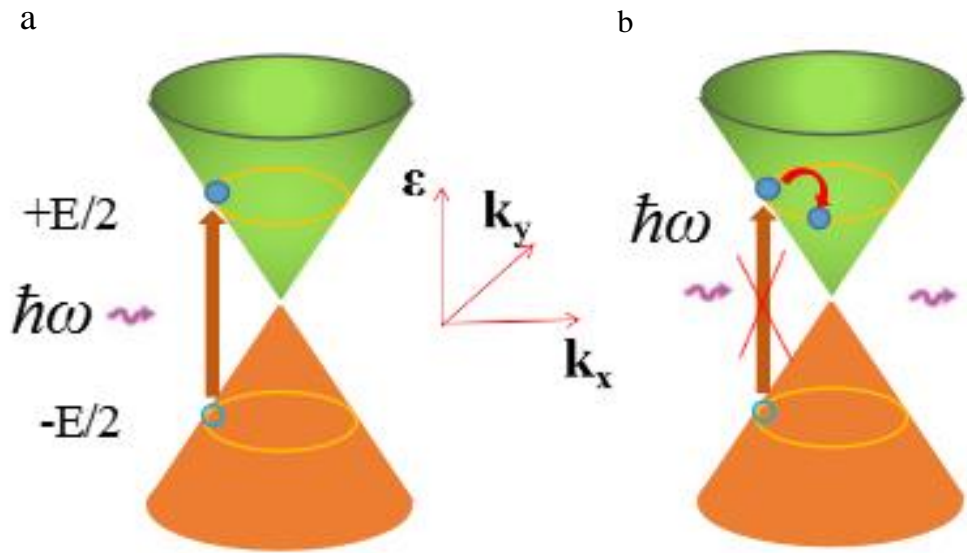

Figure 6. The schematically absorption process under different irradiations. (a) Linear absorption process under week irradiations corresponding to interband transition from the valence states to the conduction states. (b) Nonlinear saturable absorption under strong irradiations corresponding the fully occupied states blocking further absorption.